\documentclass{PoS}
\usepackage{axodraw2}
\usepackage{amsmath}
\usepackage{amssymb}
\usepackage{cite}

\title{The method of global R* and its applications}

\ShortTitle{The method of global R* and its applications}

\author{K. G. Chetyrkin\\
        University of Hamburg,GERMANY\\
        E-mail: \email{konstantin.chetyrkin@desy.de}}

\author{\speaker{G. Falcioni}\\
       Nikhef,Amsterdam,NETHERLANDS\\
       E-mail: \email{gfalcion@nikhef.nl}}

\author{F. Herzog\\
       Nikhef and VU University Amsterdam,NETHERLANDS\\
       E-mail: \email{fherzog@nikhef.nl}}

\author{J. A. M. Vermaseren\\
       Nikhef,Amsterdam,NETHERLANDS\\
       E-mail: \email{t68@nikhef.nl}}

\abstract{The global $R^*$ operation is a powerful method for
  computing renormalisation group functions. This technique, based on
  the principle of {\textit{infrared rearrangement}}, allows to
  express all the ultraviolet counterterms in terms of massless
  propagator integrals. In this talk we present the main features of
  global $R^*$ and its application to the renormalisation of QCD. By
  combining this approach with the use of the program
  Forcer for the evaluation of the relevant Feynman integrals, we renormalise for the first time QCD at five loops in covariant gauges.}

\FullConference{13th International Symposium on Radiative Corrections (Applications of Quantum Field Theory to Phenomenology)\\
         25-29 September, 2017 \\
         St. Gilgen, Austria}

\begin{document}

\section{Introduction}
Renormalisation group functions, such as beta functions and anomalous
dimensions, are crucial objects in quantum field theories, that are
necessary to define consistently Green functions and amplitudes beyond
tree level. The computation of multiloop anomalous dimensions is
therefore important both for theoretical studies and for
phenomenological applications. In this respect, a central role is
played by the renormalisation of QCD, which has a long history,
starting with the one-loop calculation and the groundbreaking
discovery of asymptotic freedom
\cite{Gross:1973id,Politzer:1973fx,Khriplovich:1969aa,tHooft:1972ikm}.
Impressive progress in this field pushed state-of-the-art calculations
to five-loop level
\cite{Caswell:1974gg,Jones:1974mm,Egorian:1978zx,Tarasov:1980au,Tarasov:1982gk,Larin:1993tp,vanRitbergen:1997va,Chetyrkin:1997dh,Vermaseren:1997fq,Chetyrkin:2004mf,Czakon:2004bu,Baikov:2014qja,Baikov:2016tgj,Luthe:2016xec,Luthe:2016ima,Herzog:2017ohr,Luthe:2017ttc,Baikov:2017ujl,Luthe:2017ttg}. One
of the main difficulties in the calculation of high order corrections
is the dramatic growth of the number and of the complexity of Feynman
diagrams, that can be tackled only with the use of highly efficient
computational methods. In this talk we describe the main features of a
particularly efficient technique to compute ultraviolet counterterms
and renormalisation constants (RCs), based on the global $R^*$
operation
\cite{Chetyrkin:2017ppe,Chetyrkin:1996sr,Chetyrkin:1996ez,Chetyrkin:1997dh},
which highly simplifies the structure of Feynman integrals. We applied
this method to the five-loop renormalisation of the QCD Lagrangian,
obtaining all the RCs with complete dependence on the gauge fixing
parameter $\xi$ \cite{Chetyrkin:2017bjc}. In this way, it was possible
to confirm and extend the results of \cite{Luthe:2017ttg}, which
include up to the linear terms in the expansion around the Feynman
gauge result, $\xi=0$.
\section{Infrared rearrangements}
\label{IRR}
The idea of {\textit{infrared rearrangement}} (IRR)
\cite{Vladimirov:1979zm} is at the basis of the method of global
$R^*$. This trick follows from the properties of dimensional
regularisation and minimal subtraction
\cite{tHooft:1972tcz,Bollini:1972ui}, which ensure that, for every
Feynman diagram $\Gamma$, its ultraviolet (UV) counterterm
$Z\left(\Gamma\right)$ is a polynomial in {\textit{all}} the masses
\cite{Collins:1974bg}. This fact implies that counterterms of
logarithmically divergent diagrams are {\textit{independent}} on
masses. As a consequence, rearranging the configuration of external
momenta or (external and/or internal) masses doesn't change the UV
counterterm of a logarithmically divergent Feynman diagram, but it can
simplify drastically the structure of the associated integral.
Notice that requiring logarithmic divergence is not restrictive,
because the superficial degree of divergence of any diagram can always
be reduced to zero by taking the appropriate number of derivatives
with respect to external momenta and masses.

The specific type of rearrangement operation is not fixed a priori and
in principle every diagram can be modified in a different way.  A
convenient choice is to reduce Feynman diagrams to ``one-mass''
tadpoles, which consists in two steps:
\begin{enumerate}
\item[i)] Selection of an external vertex attached to two
  propagators. These internal lines are modified with the introduction
  of an internal mass.
\item[ii)] Nullification of all the external momenta of the diagram.
\end{enumerate}
For example, this procedure can be used to determine the vertex
counterterm
\begin{equation}
  \begin{axopicture}(125,40)
    \Text(0,20)[lc]{$Z\Bigg($}
    \Line(15,20)(20,20)
    \Arc(33,20)(13,90,270)
    \Arc(33,20)(13,-90,90)
    \Arc(46,7)(13,90,180)
    \Line(33,38)(33,33)
    \Line(33,7)(33,33)
    \Line(46,20)(51,20)
    \Vertex(20,20){1}
    \Vertex(33,33){1}
    \Vertex(33,7){1}
    \Vertex(46,20){1}
    \Text(50,20)[lc]{$\Bigg)\,=\,Z\Bigg($}
    \Arc[double,sep=1](105,20)(13,90,270)
    \Arc(105,20)(13,-90,90)
    \Arc(118,7)(13,90,180)
    \Line(105,7)(105,33)
    \Vertex(105,7){1}
    \Vertex(105,33){1}
    \Vertex(118,20){1}
    \Vertex(92,20){1}
    \Text(120,20)[lc]{$\Bigg),$}
  \end{axopicture}
\label{eq:IRR0}
\end{equation}
where the two propagators attached to the leftmost vertex are modified
with the introduction of a mass, represented with a double line. While
the UV counterterms of the two diagrams in eq. (\ref{eq:IRR0}) are
equal to each other, the tadpole is computed much more easily using
the factorisation
\begin{equation}
  \begin{axopicture}(180,40)
    \Arc[double,sep=1](30,20)(15,90,270)
    \Arc(30,20)(15,-90,90)
    \Arc(45,5)(15,90,180)
    \Line(30,5)(30,35)
    \Vertex(30,5){1}
    \Vertex(30,35){1}
    \Vertex(15,20){1}
    \Vertex(45,20){1}
    \Text(51,20)[lc]{$=$}
    \Arc[double,sep=1](80,20)(15,90,270)
    \SetColor{Blue}
    \Arc(80,20)(15,-90,90)
    \Vertex(80,5){1}
    \Vertex(80,35){1}
    \Text(90,20)(90)[c]{\tiny $4-d$\normalsize}
    \SetColor{Black}
    \Vertex(65,20){1}
    \Text(105,20){$\cdot$}
    \Line(115,20)(125,20)
    \Arc(140,20)(15,0,360)
    \Arc(155,35)(15,180,270)
    \Line(155,20)(165,20)
    \Vertex(125,20){1}
    \Vertex(140,35){1}
    \Vertex(155,20){1}
    \Text(170,20)[lc]{,}
  \end{axopicture}
  \label{eq:omt}
\end{equation}
where the one-loop tadpole has a massless propagator raised to
non-integer power $4-d$. Note that the most complicated integral we
have to compute in eq. (\ref{eq:omt}) is a two-loop massless
propagator, while the original diagram was a three-loop one. This fact
is a particular case of a general theorem \cite{Chetyrkin:1984xa},
which shows that in principle UV counterterms of $L$-loop diagrams are
entirely determined in terms of $(L-1)$-loop massless propagators,
known also as {\textit{p-integrals}}.

One potential problem of infrared rearrangements is the generation of
spurious infrared divergences that contaminate the singularities of UV
origin. This issue occurs for example if we set to zero the internal
mass in eq. (\ref{eq:IRR0}). Of course, this rearrangement doesn't
affect the UV pole, but the tadpole integral becomes scaleless and it
vanishes in dimensional regularisation, as a consequence of the exact
cancellation of UV and IR poles. Several strategies have been adopted
to overcome the problem of IR singularities. One possibility consists
in the introduction of an internal mass, which regulates potential IR
divergences, in {\textit{all}} the propagators
\cite{Misiak:1994zw,Chetyrkin:1996vx,vanRitbergen:1997va}. After
nullification of the external momenta, this operation transforms every
$L$-loop Feynman diagram in a completely massive tadpole of the same
loop order, which in general is not factorisable in the form of
eq. (\ref{eq:omt}). This method has been applied to the five-loop
renormalisation of QCD in a series of recent works
\cite{Luthe:2016ima,Luthe:2016xec,Luthe:2017ttc}, culminating with the
calculation of all the anomalous dimensions, expanded up to the linear
order in the gauge fixing parameter $\xi$, that was presented at this
conference \cite{Luthe:2017ttg}.

A different strategy is based on the $R^*$ operation
\cite{Chetyrkin:1984xa,Chetyrkin:1982nn,Smirnov:1986me}, which
generalises the Bogoliubov $R$-operation, by subtracting both IR
and UV divergences of Feynman diagrams
\begin{equation}
  R^*(\Gamma)=\widetilde{R}\circ R(\Gamma),
  \label{eq:R*def}
\end{equation}
where the operations $R$ and $\widetilde{R}$ generate recursively UV
and IR counterterms, respectively. The power of $R^*$ is that it can
be used to cancel IR poles of rearranged Feynman integrals, thus
removing every restriction on the use of IRR. However, in practice
computations become demanding at high perturbative orders, because
each Feynman diagram generates many IR counterterms, and the
combinatorial growth of the number of diagrams makes it impossible to
proceed with this method by hand. The important step of automating the
$R^*$ operation for generic Feynman diagrams \cite{Herzog:2017bjx} was
crucial for obtaining the five-loop QCD beta function within this
method \cite{Herzog:2017ohr}.

The global $R^*$ operation \cite{Chetyrkin:1996sr,Chetyrkin:1996ez}
provides a very elegant solution to the computational issues related
to the proliferation of IR counterterms, by implementing the
subtraction of infrared singularities at the level of the whole
rearranged Green functions. In this way we avoid the
diagram-by-diagram recursive calculation of counterterms, which
results in a more efficient approach.

%
\section{Introduction to global $R^*$}
\label{globR*}
%
In this section we describe the main features of the $R^*$ operation,
by discussing the renormalisation of the ghost-gluon vertex as
illustrative example. We define the 1PI vertex
\begin{equation}
  \Gamma^{abc}_\mu(p,q)=-g_sf^{abc}\left[p_\mu\,\Gamma_p(p,q)+q_\mu\,\Gamma_q(p,q)\right],
  \label{eq:Gmabc}
\end{equation}
where $p$ is the momentum of the outgoing ghost, $q$ the momentum of
the gluon and $g_s$ is the QCD coupling constant. At tree level the
functions $\Gamma_p^{\text{tree}}=1$ and $\Gamma_q^{\text{tree}}=0$
are fixed by the Feynman rule and it is convenient to introduce the
notation
\begin{equation}
  \Gamma_p(p,q)=1+\delta\Gamma_p(p,q),\qquad
  \Gamma_q(p,q)=\delta\Gamma_q(p,q),
\end{equation}
to distinguish the tree level contribution from the loop corrections.
The renormalisation constant $Z_1^{ccg}=1+\delta Z_1^{ccg}$
satisfies the conditions 
\begin{align}
  \begin{split}
  \label{eq:verRen}  
  &K_\epsilon\left[Z_1^{ccg}\left(1+\delta\Gamma_p^B(p,q)\right)\right]=0,\\
  &K_\epsilon\left[Z_1^{ccg}\delta\Gamma_q^B(p,q)\right]=0,
  \end{split}
\end{align}
where the operator $K_\epsilon$ extracts the pole part in the Laurent
expansion in the dimensional regulator $\epsilon=\frac{4-d}{2}$ and
the superscript ``B'' indicates the use of bare lagrangian parameters
$g_0$, $\xi_0$. The first identity of eq. (\ref{eq:verRen}) determines
$Z_1^{ccg}$ order-by-order in perturbation theory
\begin{equation}
  \delta Z_1^{ccg}=-K_\epsilon\left[Z_1^{ccg}\,\delta\Gamma_p^B(p,q)\right].
  \label{eq:Z1}
\end{equation}

The first point that we want to discuss is the determination of a
global IR counterterm for the whole vertex function, which becomes IR
singular for $p,q\rightarrow 0$. In this limit IR poles cancel the UV
singularities exactly, because all the integrals become scaleless. The
key observation is that $Z_1^{ccg}$ is independent on the value of $p$
and $q$ and therefore eq. (\ref{eq:Z1}) holds also at zero momenta,
provided IR singularities are subtracted by $\widetilde{R}$
\begin{equation}
  \delta Z_1^{ccg}=-K_\epsilon\left[Z_1^{ccg}\,\widetilde{R}\left(\delta\Gamma_p^B(0,0)\right)\right].
\end{equation}
Within minimal subtraction $Z_1^{ccg}$ and $\widetilde{R}\left(\delta\Gamma_p^B(0,0)\right)$ contain only poles in $\epsilon$. This fact implies that we can drop the
operator $K_\epsilon$ from the equation above and we
derive the {\textit{IR subtracted}} vertex 
\begin{equation}
  \widetilde{R}\left(\delta\Gamma_p^B(0,0)\right)=-\frac{\delta Z_1^{ccg}}{Z_1^{ccg}}.
  \label{eq:IRct}
\end{equation}

\begin{figure}
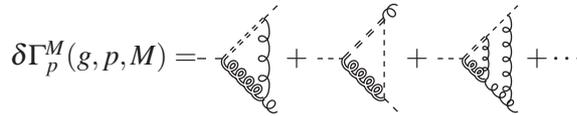

  \centering
      \begin{axopicture}(200,40)
      \Text(0,20)[l]{$\delta\Gamma_p^M(g,p,M)=$}
      \Line[dash](70,20)(80,20)
      \Line[double,sep=1,dash,dsize=2](80,20)(95,35)
      \DoubleGluon(80,20)(95,5){-2}{4}{1}
      \Gluon(95,5)(95,35){-2}{3}
      \Gluon(95,5)(100,0){-2}{1}
      \Line[dash,dsize=2](95,35)(100,40)
      \Text(103,20)[l]{$+$}
      \Line[dash,dsize=2](115,20)(125,20)
      \Line[double,sep=1,dash,dsize=3](125,20)(140,35)
      \DoubleGluon(125,20)(140,5){-2}{4}{1}
      \Line[dash,dsize=2](140,5)(140,35)
      \Line[dash,dsize=2](140,5)(145,0)
      \Gluon(140,35)(145,40){2}{1}
      \Text(148,20)[l]{$+$}
      \Line[dash,dsize=2](160,20)(170,20)
      \Line[dash,dsize=2,double,sep=1](170,20)(178,28)
      \DoubleGluon(170,20)(178,12){-1.6}{2}{1}
      \Line[dash,dsize=2](178,28)(190,40)
      \Gluon(178,12)(190,0){-1.6}{3}
      \Gluon(178,12)(178,28){-1.1}{3}
      \Gluon(186,3)(186,36.5){-1.6}{5}
      \Text(193,20)[l]{$+\cdots$}
      \end{axopicture}
      \caption{Global infrared rearrangement of the ghost-gluon vertex.}
      \label{fig:masses}
\end{figure}
The second ingredient of the procedure is the choice of a convenient
infrared rearrangement. As discussed in sec. \ref{IRR}, we want to
reduce the complexity of the calculation to $(L-1)$-loop massless
propagators. This result is achieved in global $R^*$ by applying the
rearrangement to one-mass tadpoles, which follows steps i) and ii) of
sec. \ref{IRR}, to {\textit{all}} the diagrams contributing to
$\Gamma^{abc}_\mu$. After masses have been introduced in all the
diagrams, as described in step i), we define a {\textit{globally
    rearranged}} vertex $\Gamma_p^M(p,q,M)$ as the sum of rearranged
diagrams, depicted in Fig. \ref{fig:masses}.  The genuine UV
divergence of this object is the same as $\Gamma_p(p,q)$, therefore it
is renormalised by $\delta Z_1^{ccg}$, while subdivergences are
modified. In particular eq. (\ref{eq:Z1}) becomes
\begin{equation}
  \delta Z_1^{ccg}=-K_\epsilon\left[\delta\Gamma_p^{M,B}(p,q,M)+\delta Z_1^{ccg}\cdot\delta\Gamma_p^B(p,q)\right],
\label{eq:step1}
\end{equation}
where, in the spirit of the $R$-operation, the subtraction of the
divergent subgraphs including the massive vertex is achieved by the
counterterm $\delta Z_1^{ccg}$ and the remaining reduced graph
consists of the massless vertex $\delta\Gamma_p^B(p,q)$, as shown in
the two-loop example of Fig. \ref{fig:step1}.
\begin{figure}
  \centering
    \begin{axopicture}(310,40)
      \Text(0,20)[l]{$Z\Big($}
          \Line[dash,dsize=2](15,20)(25,20)
          \Line[double,sep=1,dash,dsize=2](25,20)(40,35)
          \DoubleGluon(25,20)(40,5){-2}{4}{1}
          \Gluon(40,5)(40,35){-2}{3}
          \Gluon(40,5)(45,0){-2}{1}
          \Line[dash,dsize=2](40,35)(45,40)
          \Text(45,20)[l]{$+$}
          \Line[dash,dsize=2](55,20)(65,20)
          \Line[dash,dsize=2,double,sep=1](65,20)(75,30)
          \DoubleGluon(65,20)(75,10){-2}{2}{1}
          \Gluon(75,10)(75,30){-2}{3}
          \Line[dash,dsize=2](75,30)(83,38)
          \Line[dash,dsize=2](83,38)(85,40)
          \Gluon(75,10)(85,0){-2}{2}
          \Gluon(83,0)(83,38){-2}{4}
          \Text(89,20)[lc]{$\Big)=-K_\epsilon\Bigg[$}
          \Line[dash,dsize=2](135,20)(145,20)
          \Line[dash,dsize=2,double,sep=1](145,20)(160,35)
          \Line[dash,dsize=2](160,35)(165,40)
          \DoubleGluon(145,20)(160,5){-2}{4}{1}
          \Gluon(160,5)(165,0){-2}{1}
          \Gluon(160,5)(160,35){-2}{3}
          \Text(165,20)[lc]{$+$}
          \Line[dash,dsize=2](175,20)(185,20)
          \Line[dash,dsize=2,double,sep=1](185,20)(195,30)
          \DoubleGluon(185,20)(195,10){-2}{2}{1}
          \Gluon(195,10)(195,30){-2}{3}
          \Line[dash,dsize=2](195,30)(203,38)
          \Gluon(195,10)(205,0){-2}{2}
          \Gluon(203,0)(203,38){-2}{4}
          \Line[dash,dsize=2](203,38)(205,40)
          \Text(208,20)[lc]{$+\,Z\Big($}
          \Line[dash,dsize=2](230,20)(240,20)
          \Line[dash,dsize=2,double,sep=1](240,20)(255,35)
          \DoubleGluon(240,20)(255,5){-2}{4}{1}
          \Gluon(255,5)(255,35){-2}{3}
          \Line[dash,dsize=2](255,35)(260,40)
          \Gluon(255,5)(260,0){-2}{1}
          \Text(260,20)[lc]{$\Big)\;*$}
          \Line[dash,dsize=2](280,20)(290,20)
          \Line[dash,dsize=2](290,20)(305,35)
          \Gluon(290,20)(305,5){-2}{4}
          \Line[dash,dsize=2](305,35)(310,40)
          \Gluon(305,5)(310,0){-2}{1}
          \Text(315,20)[lc]{$\Bigg]$}
          \Gluon(305,5)(305,35){-2}{3}
    \end{axopicture}
    \caption{Pattern of divergences of the rearranged vertex $\Gamma_p^M(p,q,M)$.}
    \label{fig:step1}
\end{figure}

We proceed with step ii) of the rearrangement, namely the
nullification of external momenta in eq. (\ref{eq:step1}).  We
evaluate the leading behaviour of the massive vertex
$\delta\Gamma_p^{M,B}(p,q,M)$ in the limit $p,q\rightarrow 0$ by
applying the well-known {\textit{hard mass expansion}} (see the books
\cite{Smirnov:1994tg,Smirnov:2002pj} and references therein)
\begin{equation}
  \delta\Gamma_p^{M,B}(p,q,M){\underset{M\gg
      p,q}{\longrightarrow}}\delta\Gamma_p^{M,B}(0,0,M) +
  \delta\Gamma_p^{M,B}(0,0,M)\cdot\delta\Gamma_p^B(p,q)+{\mathcal{O}}\left(\frac{1}{M^2}\right),
  \label{eq:HME}
\end{equation}
where $\delta\Gamma_p^{M,B}(0,0,M)$ by construction is a sum of
one-mass tadpole diagrams. Finally we cancel the IR poles of the
vertex function $\delta\Gamma_p^B(p,q)$ at vanishing external momenta
by using the globally subtracted vertex
$\widetilde{R}(\delta\Gamma_p^B(0,0))$, eq. (\ref{eq:IRct}). In
conclusion, by introducing eqs.  (\ref{eq:HME}) and (\ref{eq:IRct}) in
eq. (\ref{eq:step1}) we conclude
\begin{equation}
  \delta Z_1^{ccg}=-K_\epsilon\Bigg[\frac{\delta\Gamma_p^{M,B}(0,0,M)}{Z_1^{ccg}}-\frac{\left(\delta Z_1^{ccg}\right)^2}{Z_1^{ccg}}\Bigg].
  \label{eq:Z1ccg}
\end{equation}
We are now able to determine $Z_1^{ccg}$ by expanding
eq. (\ref{eq:Z1ccg}) order-by-order in perturbation theory. The
necessary ingredients to compute $\delta Z_1^{ccg}$ at $L$ loops are
$Z_1^{ccg}$ at $(L-1)$ loops and $\delta\Gamma_p^{M,B}(0,0,M)$ at $L$
loops, that is anyway factorisable in $(L-1)$-loop massless
propagators, as in the example in eq. (\ref{eq:omt}). We calculated
these integrals up to four loops with the program \verb+Forcer+
\cite{Ruijl:2017cxj} and therefore we were able to determine the RC
$Z_1^{ccg}$ at five-loop level.

\section{The five-loop renormalisation of QCD}
We complete the renormalisation of QCD to five loops within the global
$R^*$ method introduced in sec. \ref{globR*}. Besides the ghost-gluon
vertex RC, the remaining quantities that we have to determine are the
wave function renormalisations of the ghost ($Z_3^c$), of the fermion
($Z_2$) and of the gluon ($Z_3$). All the other RCs are fixed by Ward
identities
\begin{equation}
  Z_g=\frac{Z_1^{ccg}}{Z_3^c\sqrt{Z_3}}=\frac{Z_1^{\psi\psi g}}{Z_2\sqrt{Z_3}}=\frac{Z_1^{3g}}{\left(\sqrt{Z_3}\right)^3}=\frac{\sqrt{Z_1^{4g}}}{Z_3},
\label{eq:ward}
\end{equation}
where $Z_1^i$ is the RC of the vertex $i$ and $Z_g$ is the coupling
constant renormalisation.
%

\begin{figure}
 \centering
 \begin{axopicture}(310,35)
 \Gluon(0,25)(10,25){1.8}{2}
 \GCirc(20,25){10}{0.8}
 \Gluon(30,25)(40,25){1.8}{2}
 \Text(20,25){$\mathrm{\Pi}$}
 \Text(45,25)[l]{$=$}
 \Gluon(58,25)(68,25){1.8}{2}
 \Arc(78,25)(10,0,360)
 \Gluon(88,25)(98,25){1.8}{2}
 \small
 \Text(80,0)[b]{${\mathrm{\Pi}_1}$}
 \normalsize
 \Text(103,25)[l]{$+$}
 \Gluon(116,25)(126,25){1.8}{2}
 \Arc[dash,dsize=2](136,25)(10,0,360)
 \Gluon(146,25)(156,25){1.8}{2}
 \small
 \Text(136,0)[b]{$\mathrm{\Pi}_2$}
 \normalsize
 \Text(161,25)[l]{$+$}
 \Gluon(174,25)(184,25){1.8}{2}
 \GluonCirc(194,25)(8,0){2}{10}
 \Gluon(204,25)(214,25){1.8}{2}
 \small
 \Text(194,0)[b]{$\mathrm{\Pi}_3$}
 \normalsize
 \Text(219,25)[l]{$+$}
 \Gluon(234,25)(244,25){1.8}{2}
 \GluonCirc(254,25)(8,0){1.8}{10}
 \Gluon(244,25)(264,25){1.4}{3}
 \Gluon(264,25)(274,25){1.8}{2}
 \small
 \Text(254,0)[b]{$\mathrm{\Pi}_6$}
 \normalsize
 \Text(279,25)[l]{$+\cdots$}
 \end{axopicture}
 \caption{The different contributions to the gluon self energy.}
 \label{fig:Pi}
\end{figure}
The calculation of the ghost and the fermion wave function
renormalisation follows the steps leading to eq. (\ref{eq:Z1ccg}),
described in sec. 3, and we get
\begin{align}
  \delta Z_3^c&=-K_\epsilon\Bigg\{\frac{Z_3^c}{Z_1^{ccg}}\Bigg[\Pi^B(0,M)-\frac{\delta  Z_3^c}{Z_3^c}\left(\delta\Gamma_p^{M,B}(0,0,M)+\delta Z_1^{ccg}\right)\Bigg]\Bigg\} \label{eq:Z3c},\\
  \delta Z_2&=-K_\epsilon\Bigg\{\frac{Z_2}{Z_1^{\psi\psi g}}\Bigg[\Sigma^B(0,M)-\frac{\delta  Z_2}{Z_2}\left(\delta\Lambda^{M,B}(0,0,M)+\delta Z_1^{\psi\psi g}\right)\Bigg]\Bigg\} \label{eq:Z2}.
\end{align}
Here $\Pi(0,M)$ is the ghost self-energy at zero momentum, where the
mass $M$ was introduced in the vertex of the incoming ghost by the
infrared rearrangement. Similarly $\Sigma(0,M)$ is the rearranged
fermion self-energy, while $\delta\Gamma_p^M$ and $\delta\Lambda^{M}$
are respectively the ghost-gluon vertex and the quark-gluon vertex
with masses inserted, that arise in the hard mass expansion of the
rearranged self-energies, as in eq. (\ref{eq:HME}). All the quantities
appearing in eq. (\ref{eq:Z3c}) and in eq. (\ref{eq:Z2}) are either
one-mass tadpoles, or QCD renormalisation constants, therefore we
could compute $Z_3^c$ and $Z_2$ up to five loops with the help of
\verb+Forcer+.

\begin{figure}
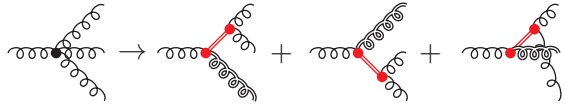

      \centering
      \begin{axopicture}(206,36)
        \Gluon(0,18)(18,18){2}{3}
        \Gluon(18,18)(36,36){2}{4}
        \Gluon(18,18)(36,18){2}{3}
        \Gluon(18,18)(36,0){2}{4}
        \Vertex(18,18){2}
        \Text(41,18)[l]{$\rightarrow$}
        \Gluon(56,18)(74,18){2}{3}
        \Red{\Line[double,sep=1](74,18)(83,27)}
        \Gluon(83,27)(92,36){2}{2}
        \Gluon(83,27)(92,18){2}{2}
        \DoubleGluon(74,18)(92,0){2}{4}{1}
        \Red{\Vertex(74,18){2}
        \Vertex(83,27){2}}
        \Text(98,18)[lc]{$+$}
        \Gluon(113,18)(131,18){2}{3}
        \DoubleGluon(131,18)(149,36){2}{4}{1}
        \Red{\Line[double,sep=1](131,18)(140,9)}
        \Gluon(140,9)(149,0){2}{2}
        \Gluon(140,9)(149,18){2}{2}
        \Red{\Vertex(131,18){2}
        \Vertex(140,9){2}}
        \Text(158,18){$+$}
        \Gluon(170,18)(188,18){2}{3}
        \Red{\Line[double,sep=1](188,18)(197,27)}
        \Gluon(197,27)(206,36){2}{2}
        \Gluon(197,27)(206,0){2}{3}
        \DoubleGluon(188,18)(206,18){2}{3}{1}
        \Red{\Vertex(188,18){2}
        \Vertex(197,27){2}}
      \end{axopicture}
      \caption{Splitting the four-point vertex into two product of two three-point vertices.}
      \label{fig:splitup}
    \end{figure}

\begin{figure}
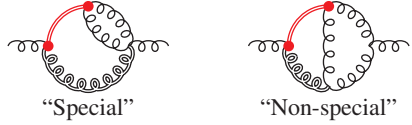

      \centering
      \begin{axopicture}(200,40)
        \Gluon(0,20)(15,20){2}{2}
        \DoubleGluonArc(30,20)(15,180,360){2}{8}{1}
        \GluonArc(30,20)(15,0,90){2}{4}
        \GluonArc(45,35)(15,180,270){2}{4}
        \Red{\DoubleArc(30,20)(15,90,180){1}
          \Vertex(15,20){2}
          \Vertex(30,35){2}}
        \Gluon(45,20)(60,20){2}{2}
        \Text(30,0)[tc]{{\footnotesize{``Special''}}}
        \Gluon(90,20)(105,20){2}{2}
        \DoubleGluonArc(120,20)(15,180,270){2}{4}{1}
        \GluonArc(120,20)(15,270,360){2}{4}
        \GluonArc(120,20)(15,0,90){2}{4}
        \Gluon(120,5)(120,35){2}{4}
        \Gluon(135,20)(150,20){2}{2}
        \Red{\DoubleArc(120,20)(15,90,180){1}
          \Vertex(105,20){2}
          \Vertex(120,35){2}}
        \Text(120,0)[tc]{{\footnotesize{``Non-special''}}}
      \end{axopicture}
      \caption{The classes of special and non-special diagrams generated by the modified four gluon vertex.}
      \label{fig:sns}
    \end{figure}

\begin{figure}
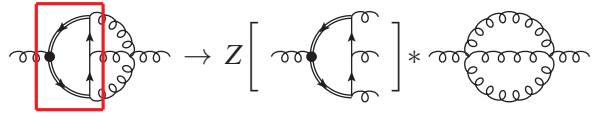

      \centering
      \begin{axopicture}(217,40)
        \Gluon(0,20)(15,20){2}{2}
        \DoubleArc[arrow,arrowscale=0.6](30,20)(15,90,180){1}
        \DoubleArc[arrow,arrowscale=0.6](30,20)(15,180,270){1}
        \Line[arrow,arrowscale=0.6](30,5)(30,20)
        \Line[arrow,arrowscale=0.6](30,20)(30,35)
        \Vertex(15,20){2}
        \GluonArc(30,20)(15,0,90){2}{4}
        \GluonArc(30,20)(15,270,360){2}{4}
        \Gluon(30,20)(45,20){2}{2}
        \Gluon(45,20)(60,20){2}{2}
        \Red{\Line[width=1.2](10,0)(35,0)
          \Line[width=1.2](10,0)(10,40)
          \Line[width=1.2](10,40)(35,40)
          \Line[width=1.2](35,0)(35,40)}
        \Text(65,20)[lc]{$\rightarrow\;Z\bigg[$}
        \Gluon(98,20)(113,20){2}{2}
        \DoubleArc[arrow,arrowscale=0.6](128,20)(15,90,180){1}
        \DoubleArc[arrow,arrowscale=0.6](128,20)(15,180,270){1}
        \Line[arrow,arrowscale=0.6](128,5)(128,20)
        \Line[arrow,arrowscale=0.6](128,20)(128,35)
        \Vertex(113,20){2}  
        \Gluon(128,35)(138,35){2}{1}
        \Gluon(128,20)(138,20){2}{1}
        \Gluon(128,5)(138,5){2}{1}
        \Text(143,20)[lc]{$\bigg]*$}
        \Gluon(157,20)(172,20){2}{2}
        \GluonArc(187,20)(15,0,180){2}{8}
        \GluonArc(187,20)(15,180,360){2}{8}
        \Gluon(172,20)(202,20){2}{4}
        \Gluon(202,20)(217,20){2}{2}
      \end{axopicture}
      \caption{Mixing between fermion and four-gluon operator in the rearranged gluon self-energy.}
      \label{fig:mixing}
    \end{figure}

The calculation of the gluon wave function renormalisation within the
global $R^*$ method is conceptually more complicated. We won't
describe here the derivation of $Z_3$, which will be given in
\cite{tobepublished}, but we will only comment on the main differences
with respect to the procedure adopted for $Z_1^{ccg}$, $Z_3^c$ and
$Z_2$. In general, infrared rearranging gluon correlators requires to
modify several types of vertices with the insertion of a mass. This
problem doesn't occur in the rearrangement of correlators with
external ghosts or fermions, where it is always possible to modify a
uniquely defined vertex in all the diagrams, because ghosts and
fermions undergo a single type of interaction. In the case of external
gluons we distinguish the contributions of the different interactions
and we rearrange them separately. As shown in Fig. \ref{fig:Pi}, for
the gluon self-energy we have\footnote{Note that the right part of
  eq. (\ref{eq:Pi}) does not include all diagrams with both external
  gluons being coupled with one and the same vertex. Such
  contributions in any case are set to zero in dimensionally regulated
  massless QCD.}
\begin{align}
  \begin{split}
    \Pi^{\mu\nu;ab}&=i\,\int d^dx \,e^{iq\cdot x}\langle 0|T\left(A^{\mu,a}(x)A^{\nu,b}(0)\right)|0\rangle_{\text{1PI}}\\
    &=i\,\sum_{i=1,2,3,6}\int d^dx \,e^{iq\cdot x}\langle 0|T\left(O_i^{\mu,a}(x)A^{\nu,b}(0)\right)|0\rangle_{\text{1PI}}\equiv\sum_{i=1,2,3,6}\Pi_i^{\mu\nu;ab},
  \label{eq:Pi}
  \end{split}
\end{align}
where $O_1^{\mu,a}$, $O_2^{\mu,a}$, $O_3^{\mu,a}$ and $O_6^{\mu,a}$
identify the operators\footnote{Colour and Lorentz indices will be
  suppressed to simplify notations.} that couple the external gluon
$A^{\mu,a}(x)$ respectively via the quark-gluon, ghost-gluon, tri- and
four-gluon interactions of QCD, {\textit{e.g.}}
$O_1^{\mu,a}=g\overline\psi\gamma^\mu T^a\psi$. Each contribution
$\Pi_i$ is rearranged by introducing a mass in $O_i$: in the case of
$O_1$, $O_2$ and $O_3$ this is straightforward. For $O_6$ we can't
directly apply steps i) and ii) of sec. \ref{IRR}, which require to
identify two propagators attached to an external leg, because here the
external gluon is connected to three lines. The solution to this
first issue is to split the four-gluon operator into three-point
vertices with the introduction of an auxiliary field, as shown in
Fig. \ref{fig:splitup}.
This procedure generates two classes of diagrams with different
structure of subdivergences, as depicted in
Fig. \ref{fig:sns}. Diagrams of the ``special'' type have
subdivergences associated to the internal vertex of the auxiliary
field, while ``non-special'' diagrams have only the singularity of
the external vertex. The properties of special and non-special
diagrams will be discussed in detail in \cite{tobepublished}.  Beyond
tree level, vertices $O_1$, $O_2$, $O_3$ and $O_6$ (both special and
non-special contributions) are mixing among each other under
renormalisation
\begin{equation}
  O^R_i = \sum_{j} z_{ij}\,O_j,
  \label{eq:mixing}
\end{equation}
where $O_i^R$ denote the renormalised vertices and $z_{ij}$ is the
renormalisation matrix. Note that the sum over $j$ is not restricted
to the QCD operators, $O_1$, $O_2$, $O_3$ and $O_6$, but it must
include {\textit{all}} the operators that cancel the UV divergences of
the vertices.
We determined the set of the required operators by analysing the
structure of the vertices at three and four loops and we identified
two new types of three-gluon interactions, named $O_4$ and $O_5$, and
six different four-gluon interactions, $O_7\dots O_{12}$ that can
appear up to three loops \cite{tobepublished}. Indeed the number of
four-gluon operators that can be constructed in a general gauge group
increases with the loop order and we limited our study to the
structures that are relevant for the renormalisation of the gluon
self-energy to five loops. In this way, renormalisation of operators
$O_i$ in eq. (\ref{eq:Pi}) generates a mixing of the different
contributions $\Pi_i$ dictated by the matrix $z_{ij}$, as shown in
Fig. \ref{fig:mixing} where $\Pi_1$ mixes into $\Pi_i$, with $i=6\dots
12$. Note however that it is not necessary to compute all the 144
matrix elements of $z_{ij}$ to renormalise eq. (\ref{eq:Pi}), because
summing over the index $i=1,2,3,6$ we have
\begin{equation}
  \sum_{i=1,2,3,6}z_{ij}=\left\{
  \begin{array}{ll}
    Z_1^{(j)}& j=1,2,3,6\\
    0& \text{otherwise}
  \end{array}
  \right.
  \label{eq:sumrule}
\end{equation}
where $Z_1^{(j)}$ are the RCs of the QCD vertices $O_j$. These
features were crucial to determine the RC $Z_3$ from one-mass tadpoles
$\Pi_i(0,M)$, that were computed with \verb+Forcer+ to five loops.

\section{Conclusions}
With the global $R^*$ operation we were able to determine all the RCs
of the QCD Lagrangian in terms of one-mass tadpoles, that are computed
at $L$-loop order, by evaluating $(L-1)$-loop massless propagators.
Crucially, the program \verb+Forcer+ allowed us to calculate these
integrals to four-loop order with high efficiency and to determine
\cite{Chetyrkin:2017bjc}
\begin{itemize}
 \item the RCs $Z_1^{ccg}$, $Z_1^{\psi\psi g}$, $Z_3^c$ and $Z_2$ to
   five loops with complete dependence on $\xi$,
 \item $Z_3$ to five loops, at linear order in $\xi$.
\end{itemize}
Because of eqs. (\ref{eq:ward}) and of the independence of $Z_g$ on
the parameter $\xi$, we derived the complete renormalisation of QCD to
five loops in general covariant gauges. Results were checked by
verifying the explicit cancellation of the linear dependence on $\xi$
in the ratio
\begin{equation}
  Z_g=\frac{Z_1^{ccg}}{Z_3^c\sqrt{Z_3}},
\end{equation}
as well as the consistency of $Z_1^{\psi\psi g}$ with the Ward
identities, eq. (\ref{eq:ward}). We compared also with the five-loop
results in Feynman gauge
\cite{Baikov:2016tgj,Herzog:2017ohr,Luthe:2017ttg} and with those
expanded up linear order in $\xi$ of \cite{Luthe:2017ttg}, finding
agreement. In Landau gauge, we verified that $Z_1^{ccg}=1$
\cite{Taylor:1971ff,Blasi:1990xz} and that $Z_3^c$, $Z_3$, $Z_2$ and
$Z_g$ agree with the results in the limit of large number of fermions
\cite{Gracey:1993ua,Gracey:1996he,Ciuchini:1999wy}.

While global $R^*$ is extremely efficient from the computational point
of view, it requires an elaborate analysis of the pattern of UV
subdivergences of the rearranged diagrams. It is interesting and
worthwhile to study $R^*$ in different contexts, where other
techniques become very demanding, such as the determination of the
anomalous dimensions of twist two operators.

\subsection*{Acknowledgement}
This work is part of the HEPGAME project, supported by the ERC
advanced grant 320651. The work of F.H. was supported by the ERC
advanced grants 320651 and 320389. The work of K.G.C. was supported by
the Deutsche Forschungsgemeinschaft through CH1479/1-1 and by the
German Federal Ministry for Education and Research BMBF through Grant
No. 05H15GUCC1.
\bibliographystyle{JHEP}
\bibliography{refs}

\providecommand{\href}[2]{#2}\begingroup\raggedright\begin{thebibliography}{10}

\bibitem{Gross:1973id}
D.J.~Gross and F.~Wilczek, \emph{{Ultraviolet Behavior of Nonabelian Gauge
  Theories}}, \href{http://dx.doi.org/10.1103/PhysRevLett.30.1343}{\emph{Phys.
  Rev. Lett.} {\bf 30} (1973) 1343}.

\bibitem{Politzer:1973fx}
H.D.~Politzer, \emph{{Reliable Perturbative Results for Strong Interactions?}},
  \href{http://dx.doi.org/10.1103/PhysRevLett.30.1346}{\emph{Phys. Rev. Lett.}
  {\bf 30} (1973) 1346}.

\bibitem{Khriplovich:1969aa}
I.B.~Khriplovich, \emph{{Green's functions in theories with non-abelian gauge
  group.}}, {\emph{Sov. J. Nucl. Phys.} {\bf 10} (1969) 235}.

\bibitem{tHooft:1972ikm}
\emph{{Proceedings, Colloquium on Renormalization of Yang-Mills Fields,
  Marseille, June 19-23, 1972}}, 1972.

\bibitem{Caswell:1974gg}
W.E.~Caswell, \emph{{Asymptotic Behavior of Nonabelian Gauge Theories to Two
  Loop Order}}, \href{http://dx.doi.org/10.1103/PhysRevLett.33.244}{\emph{Phys.
  Rev. Lett.} {\bf 33} (1974) 244}.

\bibitem{Jones:1974mm}
D.R.T.~Jones, \emph{{Two Loop Diagrams in Yang-Mills Theory}},
  \href{http://dx.doi.org/10.1016/0550-3213(74)90093-5}{\emph{Nucl. Phys.} {\bf
  B75} (1974) 531}.

\bibitem{Egorian:1978zx}
E.~Egorian and O.V.~Tarasov, \emph{{Two Loop Renormalization of the {QCD} in an
  Arbitrary Gauge}}, {\emph{Teor. Mat. Fiz.} {\bf 41} (1979) 26}.

\bibitem{Tarasov:1980au}
O.V.~Tarasov, A.A.~Vladimirov and A.{\relax Yu}.~Zharkov, \emph{{The
  Gell-Mann-Low Function of QCD in the Three Loop Approximation}},
  \href{http://dx.doi.org/10.1016/0370-2693(80)90358-5}{\emph{Phys. Lett.} {\bf
  93B} (1980) 429}.

\bibitem{Tarasov:1982gk}
O.V.~Tarasov, \emph{Anomalous dimensions of quark masses in three loop
  approximation},  \href{https://arxiv.org/abs/JINR-P2-82-900}{{\tt
  JINR-P2-82-900}}.

\bibitem{Larin:1993tp}
S.A.~Larin and J.A.M.~Vermaseren, \emph{{The Three loop QCD Beta function and
  anomalous dimensions}},
  \href{http://dx.doi.org/10.1016/0370-2693(93)91441-O}{\emph{Phys. Lett.} {\bf
  B303} (1993) 334} [\href{https://arxiv.org/abs/hep-ph/9302208}{{\tt
  hep-ph/9302208}}].

\bibitem{vanRitbergen:1997va}
T.~van~Ritbergen, J.A.M.~Vermaseren and S.A.~Larin, \emph{{The Four loop beta
  function in quantum chromodynamics}},
  \href{http://dx.doi.org/10.1016/S0370-2693(97)00370-5}{\emph{Phys. Lett.}
  {\bf B400} (1997) 379} [\href{https://arxiv.org/abs/hep-ph/9701390}{{\tt
  hep-ph/9701390}}].

\bibitem{Chetyrkin:1997dh}
K.G.~Chetyrkin, \emph{{Quark mass anomalous dimension to ${\cal
  O}(\alpha_s^4)$}},
  \href{http://dx.doi.org/10.1016/S0370-2693(97)00535-2}{\emph{Phys. Lett.}
  {\bf B404} (1997) 161} [\href{https://arxiv.org/abs/hep-ph/9703278}{{\tt
  hep-ph/9703278}}].

\bibitem{Vermaseren:1997fq}
J.A.M.~Vermaseren, S.A.~Larin and T.~van~Ritbergen, \emph{The 4-loop quark mass
  anomalous dimension and the invariant quark mass}, {\emph{Phys. Lett.} {\bf
  B405} (1997) 327} [\href{https://arxiv.org/abs/hep-ph/9703284}{{\tt
  hep-ph/9703284}}].

\bibitem{Chetyrkin:2004mf}
K.G.~Chetyrkin, \emph{{Four-loop renormalization of QCD: Full set of
  renormalization constants and anomalous dimensions}},
  \href{http://dx.doi.org/10.1016/j.nuclphysb.2005.01.011}{\emph{Nucl. Phys.}
  {\bf B710} (2005) 499} [\href{https://arxiv.org/abs/hep-ph/0405193}{{\tt
  hep-ph/0405193}}].

\bibitem{Czakon:2004bu}
M.~Czakon, \emph{{The Four-loop QCD beta-function and anomalous dimensions}},
  \href{http://dx.doi.org/10.1016/j.nuclphysb.2005.01.012}{\emph{Nucl. Phys.}
  {\bf B710} (2005) 485} [\href{https://arxiv.org/abs/hep-ph/0411261}{{\tt
  hep-ph/0411261}}].

\bibitem{Baikov:2014qja}
P.A.~Baikov, K.G.~Chetyrkin and J.H.~K{\"u}hn, \emph{{Quark Mass and Field
  Anomalous Dimensions to ${\cal O}(\alpha_s^5)$}},
  \href{http://dx.doi.org/10.1007/JHEP10(2014)076}{\emph{JHEP} {\bf 10} (2014)
  076} [\href{https://arxiv.org/abs/1402.6611}{{\tt arXiv:1402.6611}}].

\bibitem{Baikov:2016tgj}
P.A.~Baikov, K.G.~Chetyrkin and J.H.~K{\"u}hn, \emph{{Five-Loop Running of the
  QCD coupling constant}},
  \href{http://dx.doi.org/10.1103/PhysRevLett.118.082002}{\emph{Phys. Rev.
  Lett.} {\bf 118} (2017) 082002} [\href{https://arxiv.org/abs/1606.08659}{{\tt
  arXiv:1606.08659}}].

\bibitem{Luthe:2016xec}
T.~Luthe, A.~Maier, P.~Marquard and Y.~Schr{\"o}der, \emph{{Five-loop quark
  mass and field anomalous dimensions for a general gauge group}},
  \href{http://dx.doi.org/10.1007/JHEP01(2017)081}{\emph{JHEP} {\bf 01} (2017)
  081} [\href{https://arxiv.org/abs/1612.05512}{{\tt arXiv:1612.05512}}].

\bibitem{Luthe:2016ima}
T.~Luthe, A.~Maier, P.~Marquard and Y.~Schr{\"o}der, \emph{{Towards the
  five-loop Beta function for a general gauge group}},
  \href{http://dx.doi.org/10.1007/JHEP07(2016)127}{\emph{JHEP} {\bf 07} (2016)
  127} [\href{https://arxiv.org/abs/1606.08662}{{\tt arXiv:1606.08662}}].

\bibitem{Herzog:2017ohr}
F.~Herzog, B.~Ruijl, T.~Ueda, J.A.M.~Vermaseren and A.~Vogt, \emph{{The
  five-loop beta function of Yang-Mills theory with fermions}},
  \href{http://dx.doi.org/10.1007/JHEP02(2017)090}{\emph{JHEP} {\bf 02} (2017)
  090} [\href{https://arxiv.org/abs/1701.01404}{{\tt arXiv:1701.01404}}].

\bibitem{Luthe:2017ttc}
T.~Luthe, A.~Maier, P.~Marquard and Y.~Schr{\"o}der, \emph{{Complete
  renormalization of QCD at five loops}},
  \href{http://dx.doi.org/10.1007/JHEP03(2017)020}{\emph{JHEP} {\bf 03} (2017)
  020} [\href{https://arxiv.org/abs/1701.07068}{{\tt arXiv:1701.07068}}].

\bibitem{Baikov:2017ujl}
P.A.~Baikov, K.G.~Chetyrkin and J.H.~K{\"u}hn, \emph{{Five-loop fermion
  anomalous dimension for a general gauge group from four-loop massless
  propagators}}, \href{http://dx.doi.org/10.1007/JHEP04(2017)119}{\emph{JHEP}
  {\bf 04} (2017) 119} [\href{https://arxiv.org/abs/1702.01458}{{\tt
  arXiv:1702.01458}}].

\bibitem{Luthe:2017ttg}
T.~Luthe, A.~Maier, P.~Marquard and Y.~Schroder, \emph{{The five-loop Beta
  function for a general gauge group and anomalous dimensions beyond Feynman
  gauge}}, \href{http://dx.doi.org/10.1007/JHEP10(2017)166}{\emph{JHEP} {\bf
  10} (2017) 166} [\href{https://arxiv.org/abs/1709.07718}{{\tt
  arXiv:1709.07718}}].

\bibitem{Chetyrkin:2017ppe}
K.G.~Chetyrkin, \emph{{Combinatorics of $\mathbf{R}$-, $\mathbf{R^{-1}}$-, and
  $\mathbf{R^*}$-operations and asymptotic expansions of feynman integrals in
  the limit of large momenta and masses}},
  \href{https://arxiv.org/abs/1701.08627}{{\tt arXiv:1701.08627}}.

\bibitem{Chetyrkin:1996sr}
K.G.~Chetyrkin, \emph{{Correlator of the quark scalar currents and Gamma(tot)
  (H ---> hadrons) at O (alpha-s**3) in pQCD}},
  \href{http://dx.doi.org/10.1016/S0370-2693(96)01368-8}{\emph{Phys. Lett.}
  {\bf B390} (1997) 309} [\href{https://arxiv.org/abs/hep-ph/9608318}{{\tt
  hep-ph/9608318}}].

\bibitem{Chetyrkin:1996ez}
K.G.~Chetyrkin, \emph{{Corrections of order alpha-s**3 to R(had) in pQCD with
  light gluinos}},
  \href{http://dx.doi.org/10.1016/S0370-2693(96)01478-5}{\emph{Phys. Lett.}
  {\bf B391} (1997) 402} [\href{https://arxiv.org/abs/hep-ph/9608480}{{\tt
  hep-ph/9608480}}].

\bibitem{Chetyrkin:2017bjc}
K.G.~Chetyrkin, G.~Falcioni, F.~Herzog and J.A.M.~Vermaseren, \emph{{Five-loop
  renormalisation of QCD in covariant gauges}},
  \href{http://dx.doi.org/10.1007/JHEP10(2017)179}{\emph{JHEP} {\bf 10} (2017)
  179} [\href{https://arxiv.org/abs/1709.08541}{{\tt arXiv:1709.08541}}].

\bibitem{Vladimirov:1979zm}
A.A.~Vladimirov, \emph{{Method for Computing Renormalization Group Functions in
  Dimensional Renormalization Scheme}},
  \href{http://dx.doi.org/10.1007/BF01018394}{\emph{Theor. Math. Phys.} {\bf
  43} (1980) 417}.

\bibitem{tHooft:1972tcz}
G.~'t~Hooft and M.J.G.~Veltman, \emph{{Regularization and Renormalization of
  Gauge Fields}},
  \href{http://dx.doi.org/10.1016/0550-3213(72)90279-9}{\emph{Nucl. Phys.} {\bf
  B44} (1972) 189}.

\bibitem{Bollini:1972ui}
C.G.~Bollini and J.J.~Giambiagi, \emph{{Dimensional Renormalization: The Number
  of Dimensions as a Regularizing Parameter}},
  \href{http://dx.doi.org/10.1007/BF02895558}{\emph{Nuovo Cim.} {\bf B12}
  (1972) 20}.

\bibitem{Collins:1974bg}
J.C.~Collins, \emph{{Structure of Counterterms in Dimensional Regularization}},
  \href{http://dx.doi.org/10.1016/0550-3213(74)90521-5}{\emph{Nucl. Phys.} {\bf
  B80} (1974) 341}.

\bibitem{Chetyrkin:1984xa}
K.G.~Chetyrkin and V.A.~Smirnov, \emph{{R* OPERATION CORRECTED}},
  \href{http://dx.doi.org/10.1016/0370-2693(84)91291-7}{\emph{Phys. Lett.} {\bf
  144B} (1984) 419}.

\bibitem{Misiak:1994zw}
M.~Misiak and M.~M{\"u}nz, \emph{{Two loop mixing of dimension five flavor
  changing operators}},
  \href{http://dx.doi.org/10.1016/0370-2693(94)01553-O}{\emph{Phys. Lett.} {\bf
  B344} (1995) 308} [\href{https://arxiv.org/abs/hep-ph/9409454}{{\tt
  hep-ph/9409454}}].

\bibitem{Chetyrkin:1996vx}
K.G.~Chetyrkin, M.~Misiak and M.~M{\"u}nz, \emph{{Weak radiative B meson decay
  beyond leading logarithms}},
  \href{http://dx.doi.org/10.1016/S0370-2693(97)00324-9}{\emph{Phys. Lett.}
  {\bf B400} (1997) 206} [\href{https://arxiv.org/abs/hep-ph/9612313}{{\tt
  hep-ph/9612313}}].

\bibitem{Chetyrkin:1982nn}
K.G.~Chetyrkin and F.V.~Tkachov, \emph{{INFRARED R OPERATION AND ULTRAVIOLET
  COUNTERTERMS IN THE MS SCHEME}},
  \href{http://dx.doi.org/10.1016/0370-2693(82)90358-6}{\emph{Phys. Lett.} {\bf
  114B} (1982) 340}.

\bibitem{Smirnov:1986me}
V.A.~Smirnov and K.G.~Chetyrkin, \emph{{R* Operation in the Minimal Subtraction
  Scheme}}, \href{http://dx.doi.org/10.1007/BF01017902}{\emph{Theor. Math.
  Phys.} {\bf 63} (1985) 462}.

\bibitem{Herzog:2017bjx}
F.~Herzog and B.~Ruijl, \emph{{The R$^{*}$-operation for Feynman graphs with
  generic numerators}},
  \href{http://dx.doi.org/10.1007/JHEP05(2017)037}{\emph{JHEP} {\bf 05} (2017)
  037} [\href{https://arxiv.org/abs/1703.03776}{{\tt arXiv:1703.03776}}].

\bibitem{Smirnov:1994tg}
V.A.~Smirnov, \emph{{Asymptotic expansions in momenta and masses and
  calculation of Feynman diagrams}},
  \href{http://dx.doi.org/10.1142/S0217732395001617}{\emph{Mod. Phys. Lett.}
  {\bf A10} (1995) 1485} [\href{https://arxiv.org/abs/hep-th/9412063}{{\tt
  hep-th/9412063}}].

\bibitem{Smirnov:2002pj}
V.A.~Smirnov, \emph{{Applied asymptotic expansions in momenta and masses}},
  {\emph{Springer Tracts Mod. Phys.} {\bf 177} (2002) 1}.

\bibitem{Ruijl:2017cxj}
B.~Ruijl, T.~Ueda and J.A.M.~Vermaseren, \emph{{Forcer, a FORM program for the
  parametric reduction of four-loop massless propagator diagrams}},
  \href{https://arxiv.org/abs/1704.06650}{{\tt arXiv:1704.06650}}.

\bibitem{tobepublished}
K.G.~Chetyrkin, G.~Falcioni, F.~Herzog and J.~Vermaseren, \emph{{A global
  infra-red rearrangement of the ultra-violet structure of QCD}}, {\emph{to be
  published} }.

\bibitem{Taylor:1971ff}
J.C.~Taylor, \emph{{Ward Identities and Charge Renormalization of the
  Yang-Mills Field}},
  \href{http://dx.doi.org/10.1016/0550-3213(71)90297-5}{\emph{Nucl. Phys.} {\bf
  B33} (1971) 436}.

\bibitem{Blasi:1990xz}
A.~Blasi, O.~Piguet and S.P.~Sorella, \emph{{Landau gauge and finiteness}},
  \href{http://dx.doi.org/10.1016/0550-3213(91)90144-M}{\emph{Nucl. Phys.} {\bf
  B356} (1991) 154}.

\bibitem{Gracey:1993ua}
J.A.~Gracey, \emph{{Quark, gluon and ghost anomalous dimensions at O(1/N(f)) in
  quantum chromodynamics}},
  \href{http://dx.doi.org/10.1016/0370-2693(93)91803-U}{\emph{Phys. Lett.} {\bf
  B318} (1993) 177} [\href{https://arxiv.org/abs/hep-th/9310063}{{\tt
  hep-th/9310063}}].

\bibitem{Gracey:1996he}
J.A.~Gracey, \emph{{The QCD Beta function at O(1/N(f))}},
  \href{http://dx.doi.org/10.1016/0370-2693(96)00105-0}{\emph{Phys. Lett.} {\bf
  B373} (1996) 178} [\href{https://arxiv.org/abs/hep-ph/9602214}{{\tt
  hep-ph/9602214}}].

\bibitem{Ciuchini:1999wy}
M.~Ciuchini, S.E.~Derkachov, J.A.~Gracey and A.N.~Manashov, \emph{{Computation
  of quark mass anomalous dimension at O(1 / N**2(f)) in quantum
  chromodynamics}},
  \href{http://dx.doi.org/10.1016/S0550-3213(00)00209-1}{\emph{Nucl. Phys.}
  {\bf B579} (2000) 56} [\href{https://arxiv.org/abs/hep-ph/9912221}{{\tt
  hep-ph/9912221}}].

\end{thebibliography}\endgroup
\end{document}